\documentclass[apj,twocolumn]{emulateapj}

\begin{document}
\title{UNUSUAL CARBONACEOUS DUST DISTRIBUTION IN PN G095.2${+}$00.7\footnote{Based in part on data collected at Subaru Telescope, which is operated by the National Astronomical Observatory of Japan.}}
\shorttitle{Unusual Carbonaceous Dust Distribution}

\author{Ryou Ohsawa, Takashi Onaka, Itsuki Sakon, Tamami I. Mori}\affil{Department of Astronomy, Graduate School of Science, The University of Tokyo, 7-3-1 Hongo, Bunkyo-ku, Tokyo 113-0033, Japan}
\author{Takashi Miyata, Kentaro Asano}\affil{Institute of Astronomy, The University of Tokyo, 2-21-1 Osawa, Mitaka, Tokyo 181-0015, Japan}
\author{Mikako Matsuura}\affil{Department of Physics \& Astronomy, University College London, Kathleen Lonsdale Building, Gower Place, London WC1E 6BT, United Kingdom}
\author{Hidehiro Kaneda}\affil{Division of Particle and Astrophysical Sciences, G, Nagoya University, Chikusa-ku, Furo-cho, Nagoya 464-8602, Japan}

\email{ohsawa@astron.s.u-tokyo.ac.jp}

\begin{abstract}
  We investigate the polycyclic aromatic hydrocarbon features in the young Galactic planetary nebula PN\,G095.2${+}$00.7 based on mid-infrared observations. The near- to mid-infrared spectra obtained with the \textit{AKARI}/IRC and the \textit{Spitzer}/IRS show the PAH features as well as the broad emission feature at 12$\,\mu$m usually seen in proto-planetary nebulae(pPNe). The spatially resolved spectra obtained with \textit{Subaru}/COMICS suggest that the broad emission around 12$\,\mu$m is distributed in a shell-like structure, but the unidentified infrared band at 11.3$\,\mu$m is selectively enhanced at the southern part of the nebula. The variation can be explained by a difference in the amount of the UV radiation to excite PAHs, and does not necessarily require the chemical processing of dust grains and PAHs. It suggests that the UV self-extinction is important to understand the mid-infrared spectral features. We propose a mechanism which accounts for the evolutionary sequence of the mid-infrared dust features seen in a transition from pPNe to PNe.
\end{abstract}
\section{Introduction}
\label{sec:introduction}
The unidentified infrared (UIR) bands are broad and strong emission features in the infrared and widely detected both in interstellar and circumstellar environments \citep[e.g.,][]{peeters_rich_2002,acke_iso_2004}. They are thought to arise from aromatic carbonaceous materials, such as polycyclic aromatic hydrocarbons \citep[PAHs, e.g.,][]{duley_infrared_1981}. PAHs absorb UV-light, excite their vibrational modes, and re-emit in the infrared as the strong emission bands (hereafter PAH features) \citep{allamandola_interstellar_1989,puget_new_1989}. Alternatively, \citet{kwok_mixed_2011} propose amorphous organic solids with a mixed aromatic-aliphatic structure as the UIR band carriers.

Theoretical models suggest that PAHs are formed in circumstellar envelopes and ejected into the interstellar medium \citep{cau_formation_2002,cherchneff_polycyclic_1992}. \citet{kwok_subarcsecond_2002} make observations of two proto-planetary nebulae (pPNe), IRAS 07134${+}$1005 and IRAS Z02229${+}$6208, with high spatial resolution and conclude that PAHs are formed during the transitional phase from pPNe to PNe. PNe and pPNe are important objects as a site of the formation and processing of PAHs.

Mid-infrared spectroscopy is a powerful tool to investigate the formation and processing of dust grains and PAHs. Evolution of the mid-infrared spectral features is investigated by \citet{garcia-lario_transition_2003} based on \textit{Infrared Space Observatory} observations of asymptotic giant branch (AGB) stars, post-AGB stars, and PNe in the Milky Way. They show that the PAH features appear only after a broad feature around 12$\,\mu$m (hereafter the 12$\,\mu$m feature) disappears at the final phase of C-rich AGB stars. \citet{stanghellini_spitzer_2007} find that the mid-infrared spectra of PNe in the Large and Small Magellanic Clouds (MCs) show a similar spectral transition along with the nebula size. Such a transition is not investigated among Galactic PNe because the 12$\,\mu$m feature is rarely detected in Galactic PNe \citep{bernard-salas_unusual_2009}. There is no convincing explanation on the evolutionary sequence of carbonaceous dust features. To understand how PAHs are formed and processed, it is important to identify the origin of the spectral transition from pPNe to PNe.

PNe and pPNe consist of several distinct physical environments such as a hot central star, ionized and photo dissociation regions, and a dust envelope. A study of the dust features associated with each environment is important for the understanding of the evolution of dust grains and PAHs in these objects. \citet{waters_oxygen-rich_1998} indicate the presence of extended PAH emissions together with the crystalline silicate emission in the central torus of a bipolar pPN, the Red\,Rectangle, which is confirmed by high spatial resolution spectroscopy by \citet{miyata_sub-arcsecond_2004}. \citet{goto_spatially_2003} find a systematic variation in the PAH features at 3.3 and 3.4$\,\micron$ with the distance from the central star in the pPN IRAS\,22272+5435, suggesting thermal processing on dust grains by stellar radiations. The spatially resolved spectroscopy is a promising method to identify the origin of the variation in the infrared spectrum.

In this paper, we present the mid-infrared spectra and image of PN\,G095.2${+}$00.4 (hereafter G095), investigate the spatial distribution of the PAH features in the circumstellar nebula, and propose a scenario to account for the spectral transition presented by \citet{garcia-lario_transition_2003} and \citet{stanghellini_spitzer_2007}. We describe the observations in Section \ref{sec:observations}. Results are given in Section \ref{sec:results}. The dust features of G095 and their distributions are discussed in Section \ref{sec:discussion}. Section \ref{sec:conclusions} summarizes the conclusions.

\section{Observations}
\subsection{Target Information}
G095 is a Galactic PN at ($\alpha$,$\delta$) ${=}$ ($21:31:50.18$, ${+}52:33:51.6$). Its distance, radio angular radius, and [\ion{O}{3}]$\lambda 5007$ expansion velocity are estimated to be 2.3$\,$kpc \citep{cahn_catalogue_1992}, $1''.3$ \citep{aaquist_six_1990}, and $16\,{\rm km\,s^{-1}}$ \citep{robinson_measurements_1982}, respectively. The estimated dynamical age of the object is about 900 year, indicating that G095 is young.

\label{sec:observations}
\begin{table*}[tb]
\centering
\caption{Summary of Observations}
\begin{tabular}{ccccc}
  \hline \hline
  Instrument
  & \textit{AKARI}/IRC & \textit{Spitzer}/IRS
  & \textit{Subaru}/COMICS & \textit{Subaru}/COMICS \\
  Obs. date
  & 2008-12-19 & 2008-08-06 & 2011-08-31 & 2011-08-31 \\
  Observation
  & Spectroscopy & Spectroscopy & Spectroscopy & Imaging \\
  Grism/filter
  & NG~($R \simeq 100$) & SL~($R \simeq 100$)
  & NL~($R \simeq 250$) & N11.7~($R \simeq 10$) \\
  Waverange
  & 2.5--5.0$\,\mu$m & 5.5--14$\,\mu$m
  & 8.0--13.5$\,\mu$m & 11.1--12.4$\,\mu$m \\
  Slit width
  & $1'$ & $3''.7$ & $0''.26$ & \nodata \\
  Exp. time
  & 799$\,$s & 12.58$\,$s & 600$\,$s & 185$\,$s \\
  Standard star
  & \nodata & \nodata & HD\,197989\,(\ion{K0}{3}) 
  & HD\,197989\,(\ion{K0}{3}) \\
  \hline \hline  
\end{tabular}

  \label{tab:observations}
\end{table*}

\begin{figure*}[tp]
  \centering
  \epsscale{1.0}\plotone{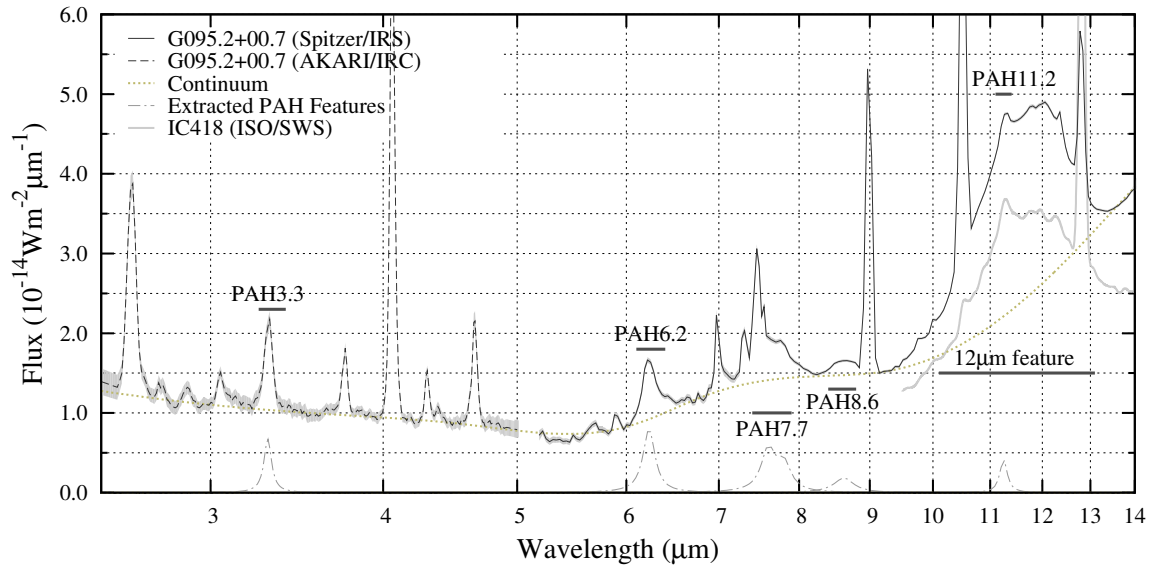}
  \caption{Observed spectrum of G095 (the solid and dashed lines). The errors are shown by the grayed regions. The dotted line indicates the continuum emission approximated by spline functions. The dash-dotted lines show the extracted PAH features. The normalized 10--14$\,\mu$m spectrum of IC\,418 is also displayed by the gray thick line.}
  \label{fig:satellitespec}
\end{figure*}

\subsection{\textit{AKARI} and \textit{Spitzer} Observations}
\label{sec:akari-spitz-observ}
The near-infrared spectrum (2.5--5.0$\,\mu$m in Figure~\ref{fig:satellitespec}) of G095 is obtained with the \textit{AKARI}/IRC \citep{onaka_infrared_2007}. Table \ref{tab:observations} summarizes details of the InfraRed Camera (IRC) observation (Pointing ID: 3460033, PI: T. Onaka). G095 is unresolved with the IRC (FWHM ${\sim}4''.7$). We use the $1'{\times}1'$ window for slitless spectroscopy to collect all the flux from the object. The data reduction is performed with the IRC spectroscopy official toolkit for phase 3 data (version 20111121).

The mid-infrared spectrum (5--14$\,\mu$m in Figure~\ref{fig:satellitespec}) is retrieved from the \textit{Spitzer} archive (Program ID: 50261, PI: L. Stanghellini). The spectrum is obtained with the short low (SL) mode of the Infrared Spectrograph \citep[IRS;][]{houck_infrared_2004}. Details of the IRS observation are also summarized in Table \ref{tab:observations}. The IRS spectrum smoothly connects with the IRC spectrum without any scaling, suggesting that almost all the flux from the object falls within the IRS slit.

\subsection{\textit{Subaru}/COMICS observations}
\label{sec:subar-observ}
\begin{figure}[tbp]
  \centering
  \epsscale{1.0}\plotone{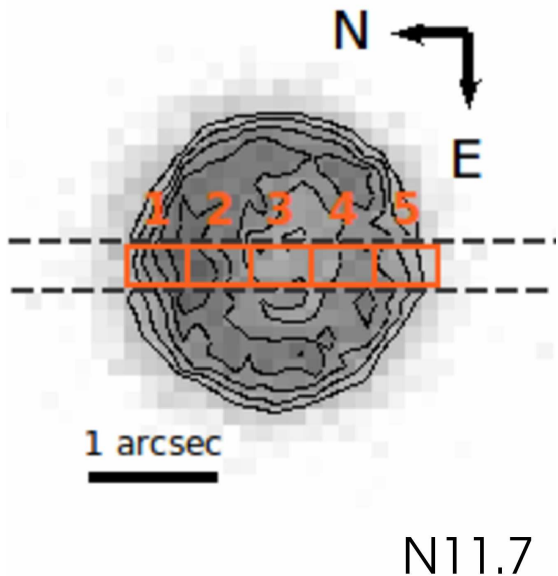}
  \caption{N11.7 image obtained with the COMICS. The thin lines show the contours from 50\% to 100\% of the peak. The dashed lines show the slit position in spectroscopy. The orange rectangles show the region where we extract the spectra.}
  \label{fig:comicsimage}
\end{figure}

We obtained the N11.7 (11.7$\,\micron$) broad-band image and the $N$-band low-resolution spectra of G095 with the \textit{Subaru}/COMICS \citep{kataza_comics:_2000} as part of the open use program (Prop. ID: S11B-017, PI: Ohsawa). Details of the observation are summarized in Table \ref{tab:observations}. The data reduction is performed in the standard manner for mid-infrared ground-based observations, including dark subtractions, flat fielding, and background subtraction from different chopping positions. The chopping throw was set as 20$''$ in the north--south direction. Figure~\ref{fig:comicsimage} shows the image of G095 taken at N11.7. The pixel scales are $0''.13$ and $0''.16$ for imaging and spectroscopy, respectively. The FWHM of the PSF is about $0''.39$. The slit position in the spectroscopy is displayed by the dashed lines in Figure~\ref{fig:comicsimage}. The five rectangles with numbers indicate the regions where we extract spectra. Region 3 corresponds to the geometric center of the nebula. We use the template spectrum of HD\,197989 provided by \citet{cohen_spectral_1999} for the calibration.

\begin{figure*}[btp]
  \centering
  \epsscale{1.0}\plotone{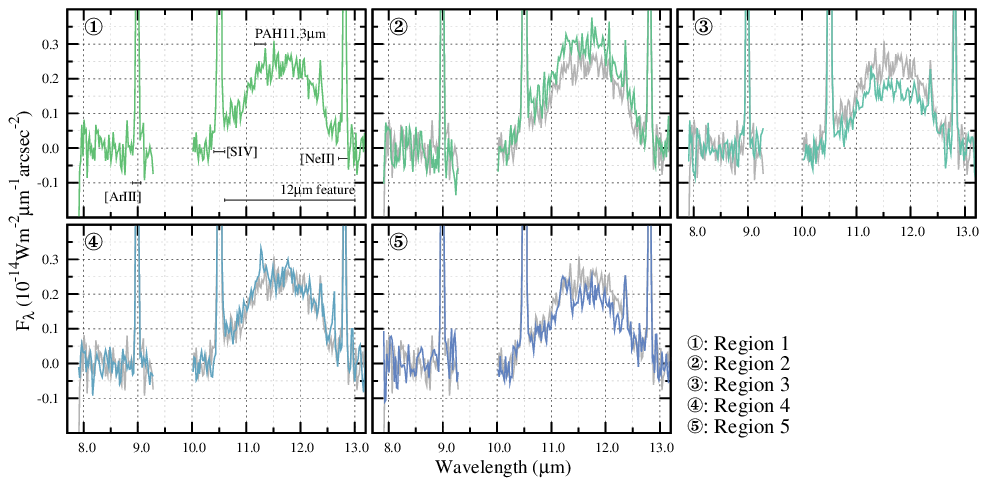}  
  \caption{Continuum-subtracted spectra obtained with the COMICS. The spectrum of Region 1 is superimposed by the gray line for comparison in other regions' spectra. The spectral range affected by terrestrial atmospheric absorption is not displayed.}
  \label{fig:comicsspectra}
\end{figure*}

\section{Results}
\label{sec:results}
\subsection{\textit{AKARI} and \textit{Spitzer} Spectra}
\label{sec:results-from-akari-spitzer}
We can see the PAH features at 3.3, 6.2, 7.7, 8.6, and 11.2$\,\mu$m in Figure~\ref{fig:satellitespec}. The IRC spectrum show the eminent PAH feature at 3.3$\,\mu$m, while the aliphatic feature at 3.4$\,\mu$m is not detected. In the IRS spectrum, the PAH features at 6.2, 7.7, 8.6, and 11.3$\,\mu$m are detected. \citet{stanghellini_nature_2012} reported that G095 does not show the PAH features, but we claim that the PAH features are certainly detected in both IRC and IRS spectra.

The intensities of the PAH features are estimated by spectral fitting with Lorentzian functions. The continuum is approximated by spline functions with anchoring points at 3.0, 3.5, 2.4, 6.0, 6.8, 9.3, 13.6, and 14.0$\,\mu$m. The continuum for the 11.3$\,\mu$m feature is approximated by spline interpolation of the adjacent regions. The flux density around 3.29$\,\mu$m also includes the contribution form Pfund-$\delta$. It is calculated from the intensity of Brackett-$\alpha$ at 4.05$\,\mu$m assuming the Case-B condition with the electron density of $10^6\,{\rm cm^{-3}}$ and the electron temperature of $10,000\,{\rm K}$ \citep{storey_recombination_1995}. The estimated intensities are $0.35 \pm 0.02$, $1.88 \pm 0.04$, $2.95 \pm 0.06$, $1.01 \pm 0.04$, and $1.06 \pm 0.04$ in units of ${\rm 10^{-15}W\,m^{-2}}$ for the PAH features at 3.3, 6.2, 7.7, 8.6, and 11.3$\,\mu$m, respectively.

Another eminent spectral feature in Figure~\ref{fig:satellitespec} is a broad feature from 10 to 12$\,\mu$m. We measure the intensity of the broad feature by integrating over the continuum and subtracting the contribution from [\ion{S}{4}] at 10.51$\,\mu$m, the 11.3$\,\mu$m PAH feature, Humphreys-$\alpha$ at 12.4$\,\mu$m and [\ion{Ne}{2}] at 12.81$\,\mu$m. The estimated intensity is $(4.90 \pm 0.01){\times}{\rm 10^{-14}W\,m^{-2}}$.

\subsection{\textit{Subaru}/COMICS image and spectra}
\label{sec:results-subaru-comics}
Figure~\ref{fig:comicsimage} shows the nebulosity of G095. It has a radius of about $1''.2$, being consistent with the radio angular radius \citep{aaquist_six_1990}.  According to Figure~\ref{fig:satellitespec}, the flux around 11.7$\,\mu$m is dominated by the broad feature around 12$\,\mu$m and the underlying continuum emission. Thus, the N11.7 image represents the distribution of the circumstellar dust envelope, indicating a shell-like structure.

The 8.4$\,$GHz radio observation \citep{kwok_vla_1993} shows a double shell structure, the diameter of which is $1''.1$ and $1''.9$, respectively. We investigate the spatial profile of the fine-structure lines in the two-dimensional spectrum of G095, but the double shell structure is not confirmed possibly due to an insufficient spatial resolution (FWHM$\sim$0$''$.48). The size of the envelope shell in the N11.7 image suggests that the N11.7 emission mainly comes from the outer shell with some contribution from the inner shell.

We extract the spectra at the position of the rectangles in Figure~\ref{fig:comicsimage}.  While the spectra of Regions 2, 3, and 4 include the emission from the inner and outer shell, the spectra of Regions 1 and 5 are dominated by the emission from the outer shell. The continuum-subtracted spectra are presented in Figure~\ref{fig:comicsspectra}. The underlying continua are approximated by spline functions with anchoring points at 8.1, 8.5, 10.1, and 13.0$\,\mu$m. 

The 11.3$\,\mu$m PAH feature is detected, while the 8.6$\,\mu$m PAH feature is not detected in any region due to the low signal-to-noise ratios at short wavelengths. We measure the brightness of the 11.3$\,\mu$m feature by spectral fitting. The continuum for the 11.3$\,\mu$m feature is estimated by spline-interpolation of the adjacent regions. The brightness of the broad feature around 12$\,\mu$m is estimated by integrating the flux from 10 to 13$\,\mu$m over the underlying continuum and subtracting the contribution from [\ion{S}{4}], the 11.3$\,\mu$m PAH feature, the Humphreys-$\alpha$, and [\ion{Ne}{2}].

\begin{figure*}[tbp]
  \centering
  \epsscale{0.95}\plotone{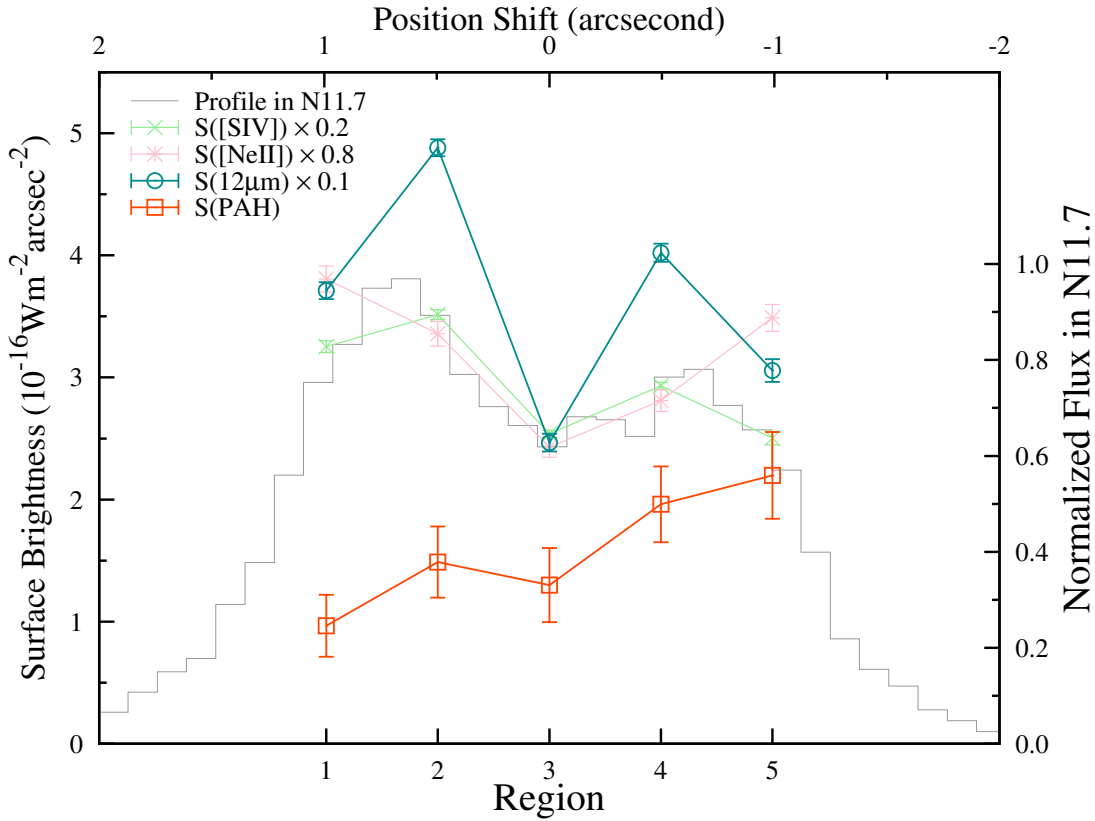}
  \caption{Intensities of the 11.3$\,\mu$m PAH feature, 12$\,\mu$m feature, [\ion{S}{4}] and [\ion{Ne}{2}] at the different regions. The thin gray line shows the profile of the N11.7 image along with the slit.}
  \label{fig:gradient}
\end{figure*}

Figure~\ref{fig:gradient} shows the surface brightness of the emission features in the COMICS spectra along the slit. The profile of the N11.7 image is also shown for comparison. The profile of the broad feature around 12$\,\mu$m shows the same spatial variation of the N11.7 image. However, the surface brightness of the 11.3$\,\mu$m PAH feature increases from Regions 1 to 5 (from north to south) and the intensity of the PAH feature in Region 5 is twice as strong as that in Region 1. Figure~\ref{fig:gradient} also plots the surface brightnesses of the fine-structure lines, [\ion{S}{4}]\,(34.79$\,$eV) and [\ion{Ne}{2}]\,(21.56$\,$eV). The [\ion{S}{4}] emission has peaks at Regions 2 and 3, while the [\ion{Ne}{2}] is strong at Regions 1 and 5.

\section{Discussion}
\label{sec:discussion}

\subsection{The PAH and 12$\,\mu$m Features}
\label{sec:pah-12-mum-features}
The intensity ratio of the PAH features are often used to diagnose the ionization and excitation conditions of PAHs. The 6.2 to 11.3$\,\mu$m and 7.7 to 11.3$\,\mu$m band ratios are about 1.8 and 2.8, respectively. These ratios become large when PAHs are ionized \citep{schutte_theoretical_1993}. The 6.2 to 11.3$\,\mu$m and the 7.7 to 11.3$\,\mu$m band ratios of typical Galactic PNe range from 0.5 to 3.5 and from 1.0 to 6.5, respectively \citep{bernard-salas_unusual_2009}. These ratios suggest that the ionization of PAHs in G095 is intermediate among Galactic PNe. 

The 11.3 to 3.3$\,\mu$m band ratio is used to estimate the excitation of PAHs. Since the 3.3$\,\mu$m feature requires a high excitation of PAHs, the band ratio becomes small when the excitation of PAHs becomes high. The ratio of G095 is about 3.0, while \citet{hony_ch_2001} reported that the typical 11.3 to 3.3$\,\mu$m band ratio ranges from 3 to 4. The small 11.3 to 3.3$\,\mu$m band ratio of G095 suggests that the PAH excitation in G095 is high. It seems consistent with the hard radiation field of PNe. On the other hand, the destruction of small PAHs increases the 11.3 to 3.3$\,\mu$m ratio as discussed in \citet{mori_observations_2012}. The small ratio of G095 also suggests that PAHs in G095 are still intact in spite of the hard radiation fields.

To investigate the carrier of the broad feature around 12$\,\mu$m, we plot the spectrum of IC\,418, which is a PN with the 12$\,\mu$m feature \citep{volk_iras_1990}, in Figure~\ref{fig:satellitespec} by the thick gray line. The broad features of G095 and IC\,418 show almost the same spectral profiles. \citet{bernard-salas_unusual_2009} also investigate the 12$\,\mu$m feature of PNe in MCs. The peak position and wavelength range of the broad feature of G095 are consistent with those of the 12$\,\mu$m feature reported in \citet{bernard-salas_unusual_2009}. Thus, we conclude that the broad feature in the IRS spectrum has the same origin of the 12$\,\mu$m features discussed in \citet{garcia-lario_transition_2003}, \citet{bernard-salas_unusual_2009}, and \citet{stanghellini_spitzer_2007}.

The 12$\,\mu$m feature is generally attributed to silicon carbide (SiC) grains \citep[e.g.,][]{speck_silicon_2009,bernard-salas_unusual_2009}. Alternatively, \citet{stanghellini_nature_2012} tentatively attribute the broad feature around 12$\,\mu$m to aliphatic carbons. If the 12$\,\mu$m feature of G095 arises from aliphatic carbons, we expect the aliphatic feature at 3.4$\,\mu$m. However, the IRC spectrum of G095 does not show any feature around 3.4$\,\mu$m. It does not support the aliphatic origin of the 12$\,\mu$m feature of G095. Thus, we conclude that the 12$\,\mu$m feature of G095 is attributable to SiC grains.

The 12$\,\mu$m feature is seldom detected in Galactic PNe. PNe in MCs show the 12$\,\mu$m feature more frequently than Galactic PNe \citep{bernard-salas_unusual_2009,stanghellini_spitzer_2007}. \citet{stanghellini_nature_2012} point out that the typical dust temperature of carbon-rich PNe in LMC is higher than that of Galactic carbon-rich PNe. Frequent detection of the 12$\,\mu$m feature in PNe in MCs is attributable to their high dust temperature. We expect that the dust temperature in G095 is sufficiently high to show the 12$\,\mu$m feature based on the small size of G095. G095 has the blue \textit{IRAS} 25 to 60$\,\mu$m color of about 2.0, while the ratio is typically about 0.7 for Galactic PNe \citep{acker_strasbourg_1992}. It also supports the high dust temperature of G095.

\subsection{Spatial Distribution of the Dust Features}
\label{sec:spatial-variation-dust}
The most simple explanation of the difference between Regions 1 and 5 is that the amount of PAHs is different between the northern and southern parts of G095. It may result from efficient PAH formation in the southern part, or PAH destruction in the northern part. However, there is no observation to support the scenario. The strength of the 12$\,\mu$m feature and underlying continuum emissions, both of which can be attributed to carbonaceous dust grains, do not show a large difference between the northern and southern part of G095. It suggests that the formation of carbonaceous grains does not differ between the northern and southern regions. If the emission from ionized gas is enhanced in the northern region, there can be PAH destruction. However, the 6$\,$cm free--free emission in the northern part is as strong as that in the southern region \citep{aaquist_six_1990}. The [\ion{S}{4}] to [\ion{Ne}{2}] intensity ratio does not show a significant difference between the northern and southern regions. In addition, the 11.3 to 3.3$\,\mu$m PAH band ratio does not support PAH destruction in G095. Although we cannot completely rule out the possibility that the southern region has a large amount of PAHs, there is no observational evidence for it.

We propose that the UV absorption within the dust envelope can explain the difference in the strength of the PAH feature between Regions 1 and 5. Since PAHs require the UV photons to be excited, a different amount of UV photons could explain the different strength of the PAH feature. Assuming that PAHs coexist with other dust grains in the nebula, UV radiation can be absorbed by dust grains before it excites PAHs. The amount of the UV photons to excite PAHs is estimated by $\exp(-\tau_{\rm UV})$, where $\tau_{\rm UV}$ is the optical depth to the central star at the UV wavelengths. The dust column density in the northern region is suggested to be higher by about 10\% than that in the southern region from the COMICS observation. Assuming that $\tau_{\rm UV}$ in the northern region is 10\% larger than in the southern region, the amount of the UV photon in the southern region can be twice larger than in the northern region when $\tau_{\rm UV} \sim 7$. Thus, we expect that G095 is optically thick in the UV. The color index of G095 is $\bv{\simeq}1.5$ \citep{abazajian_seventh_2009}. Assuming $R_V{=}3.1$ and the intrinsic color ${\sim}0$, the amount of the extinction without the interstellar extinction ($A_V{\sim}0.31$ from \citet{dobashi_atlas_2011}) is roughly estimated as $A_V{\sim}4.34$, corresponding to $A_{\rm 0.15\,\mu m}{\simeq}11.5$ or $\tau_{\rm 0.15\,\mu m}{\simeq}10.6$ \citep{mathis_interstellar_1990}. The estimated optical depth is sufficiently large to quantitatively explain the difference in the strength of the 11.3$\,\mu$m PAH feature between Regions 1 and 5. Thus, a small asymmetry in the amount of dust ($\sim$10\%) can result in a large difference in the intensity of the PAH feature.

The present results suggest that we have to consider the UV absorption within the nebula to investigate the PAH features in young PNe, such as G095. We speculate that part of the mid-infrared spectral sequence from pPNe to PNe can be explained in terms of the UV absorption even if there is no assumption of dust processing. In the early phase of the sequence, the circumstellar nebula is small and dense so that it can be optically thick in the UV. Nearly all the UV photons are absorbed at the inner surface of the dust envelope and PAHs are not excited efficiently. The small size of the nebula leads to dust temperature high enough to produce the 12$\,\mu$m feature. As the nebula expands along the evolution, the optical depth in the UV and the dust temperature decrease. PAHs start to be excited by the UV photons and the PAH features are efficiently emitted. The drop in the dust temperature results in a decrease of the 12$\,\mu$m feature. Consequently, the mid-infrared spectrum becomes dominated by the strong PAH emission. The advent of the PAH features and the disappearance of the 12$\,\mu$m feature of the mid-infrared spectral sequence can be interpreted in terms of the UV absorption within the nebula and the nebular expansion.

\section{Summary and Conclusions}
\label{sec:conclusions}
We investigate the PAH features of G095 based on the infrared images and spectra. We detect the PAH features and the 12$\,\mu$m feature in the IRC and IRS spectra. We obtain the mid-infrared image and spatially resolved spectra in the $N$-band with the \textit{Subaru}/COMICS. The surface brightness of the 11.3$\,\mu$m PAH feature in the southern region is twice stronger than in the northern region, while the 11.7$\,\mu$m image and the 12$\,\mu$m feature do not show a large difference between the northern and southern parts of the nebula. We propose that the difference in the strength of the PAH feature is caused by the UV absorption of the dust envelope rather than the chemical processing of carbonaceous dust. 

The present results suggest that the UV absorption within the nebula is needed to be taken into account in the investigation of the mid-infrared spectrum of pPNe and young PNe. We propose the scenario to account for the mid-infrared spectral transition seen in the evolution from pPNe to PNe in terms of the UV absorption and the nebular expansion without assuming any dust processing.

The mid-infrared spectral evolution in a transition from pPNe to PNe involves the evolution of nebulosity as well as the processing of dust grains. We propose that the change in the UV absorption within the nebula can significantly affect the mid-infrared spectrum. Further studies will confirm that the proposed scenario is applicable to other pPNe and young PNe.

\acknowledgments
This result is based on data collected at Subaru Telescope, which is operated by the National Astronomical Observatory of Japan. It is in part based on observations with \textit{AKARI}, a JAXA project with the participation of ESA and on archival data obtained with the \textit{Spitzer Space Telescope}, which is operated by the Jet Propulsion Laboratory, California Institute of Technology under a contract with NASA. We fully appreciate all the people who worked in the operation and maintenance of those instruments. This work is supported in part by a Grant-in-Aid for Scientific Research (23244021) by the Japan Society of Promotion of Science (JSPS).

Facilities: 
\facility{AKARI(IRC)},
\facility{Spitzer(IRS)},
\facility{Subaru(COMICS)}
\acknowledgements

\bibliographystyle{apj}

\end{document}